# A Normative Dual-value Theory for Bitcoin and other Cryptocurrencies

Zhiyong Tu
Associate Professor in Economics
Peking University HSBC Business School
University Town, Shenzhen, China 518055
zytu@phbs.pku.edu.cn

Lan Ju
Associate Professor in Finance
Peking University HSBC Business School
University Town, Shenzhen, China 518055
julan@phbs.pku.edu.cn

April, 2019

Abstract

Bitcoin as well as other cryptocurrencies are all plagued by the impact from bifurcation. Since the marginal cost of bifurcation is theoretically zero, it causes the coin holders to doubt on the existence of the coin's intrinsic value. This paper suggests a normative dual-value theory to assess the fundamental value of Bitcoin. We draw on the experience from the art market, where similar replication problems are prevalent. The idea is to decompose the total value of a cryptocurrency into two parts: one is its art value and the other is its use value. The tradeoff between these two values is also analyzed, which enlightens our proposal of an image coin for Bitcoin so as to elevate its use value without sacrificing its art value. To show the general validity of the dual-value theory, we also apply it to evaluate the prospects of four major cryptocurrencies. We find this framework is helpful for both the investors and the exchanges to examine a new coin's value when it first appears in the market.

# 1. Introduction

The Bitcoin market crashed by 72% over the year of 2018.[1] The investment capital associated to the whole ecology of the cryptocurrencies suffered huge losses. Most importantly, people's belief in Bitcoin was shaken.

Satoshi Nakamoto, the creator of Bitcoin, claimed that Bitcoin could serve as an alternative to the sovereign money system. The message left by Nakamoto in the first block of Bitcoin expressed his dissent on the abuse of the central banking scheme during the 2008 financial crisis.[2] The concept of Bitcoin quickly obtained popularity among believers and its price soared from worthless to nearly $20,000 in December, 2017.

The success of Bitcoin leads to the emergence of a large number of competing counterparts, such as Ethereum, Litecoin and Bitcoin Cash, etc. Currently the CoinMarketCap lists more than 2000 cryptocurrencies, among which Bitcion still takes more than half of the total market capitalization.

As the cryptocurrencies form a completely new financial ecosystem, they have aroused great academic interests in both science and social science. Many traditional research topics in finance are also projected into the cryptocurrency market. For example, some modeled the price trends of Bitcoin (Cheung et al., 2015; Cagli, 2018), some studied the volatility patterns of cryptocurrencies (Urquhart, 2016; Charles and Darne, 2019), some investigated the market efficiency problem (Nadarajah and Chu, 2017; Bouri et al., 2019), and others explored the risk contagion issue (Yi et al., 2018; Katsiampa, 2018).

In spite of the fast-growing literature, the critical question that whether Bitcoin does have an intrinsic value is still not fully addressed. Previous studies normally approach this problem empirically (e.g. Cheah and Fry, 2015), while this paper will try to propose a normative theory on the Bitcoin value.

Unlike the sovereign money, Bitcoin is not backed by any entity. We believe the increasing doubt on the existence of the fundamental value of Bitcoin, as well as those of other cryptocurrencies, is the key to explain the recent huge fluctuation of the cryptocurrency markets. When passion returns to reason, this question cannot be bypassed if the cryptocurrency markets are to survive into the future.

The supporters of Bitcoin believe that the Bitcoin technology provides a well-functioning decentralized transaction system. Its value should not be doubted as it has already stood up the test of past ten years' practice. However, the recent market down turn echoes many investors' shaken belief.

---

[1] According to CoinMarketCap, the closing price of Bitcoin was $13657 on Jan 1, 2018, and it dropped to $3844 on Jan 1, 2019.
[2] The message left by Nakamoto in the first block of Bitcoin's blockchain is: The Times 03/Jan/2009 Chancellor on the brink of second bailout for banks.

In order to analyze the Bitcoin value, what factors shall we consider then? We suggest the following two: the first is the degree to which Bitcoin can be integrated into our everyday life; and the second is the extent to which Bitcoin can be guarded against the impact from its own bifurcation.[3] We hold that these are the two pillars of Bitcoin value, both of which are currently under serious challenges.

Normally, the more often Bitcoin is used in reality, the higher value it will reach. The less it is threatened or diluted by the bifurcation, the greater value it can obtain. We call the former Bitcoin's *use value* and the latter its *art value*. This paper will propose a normative theory for the value of Bitcoin based on these two values. The theory provides a conceptual framework to look at Bitcoin's fundamental, which leads to a policy suggestion that may lift Bitcoin's value in the future. Our theory can be applied to the other cryptocurrencies in a similar fashion.

In short, the theory proposes that the Bitcoin value can be written as the product of its art value and its use value. Each argument is scaled up or down by the other argument, while at the same time there also exists certain inherent tradeoff between these two arguments. For example, in order to preserve the art value, Bitcoin shall not modify its original algorithm. But its rapidly increasing usage is confined by the old system design, which calls for some updating. To resolve this conflict between the art and use values of Bitcoin, we suggest a scheme of a pegged image coin as a possible solution.

Section 2 will lay out the theory. Section 3 and 4 elaborate on the art and use values of Bitcoin respectively. Section 5 examines the conflict between these two values. Section 6 applies the theory to several major cryptocurrencies in the market and evaluates their prospects. Section 7 concludes.

## 2. The normative dual-value theory for Bitcoin

Most empirical studies investigate the functioning of Bitcoin system by assuming that it has an intrinsic value. For example, Ju et al. (2016) provide evidence that people use Bitcoin as a venue for capital flight, which naturally assumes that Bitcoin can carry value.

However, what anchors the intrinsic value of Bitcoin remains ambiguous. Especially, in a purely laissez-faire environment for Bitcoin, the marginal cost of replication of the Bitcoin platform (code) is theoretically zero. Because it is open-sourced, and no legal entity protects its intellectual property. Economic theory will generally predict that its price will converge to its marginal cost, which is zero.

Then a theory that underlies the intrinsic value of Bitcoin will have to move away from the traditional economic doctrine, and rest upon the notion of belief/consensus rather than the economic cost of replication. That motivates our proposal that a necessary dimension of Bitcoin's value shall resemble the value of an artwork, i.e., an artwork can have a great value even though it can be easily reproduced.

---

[3] A bifurcation is a hard fork (without user consensus) from the blockchain of the original coin, which normally leads to a new coin.

On the other hand, Bitcoin is not created to be hanged on the wall like paintings; rather its aim from its creator Nakamoto is to act as an alternative transaction media to challenge the sovereign money. To fulfill this goal, it must be used in our economic life, which represents its use value dimension.

By the above logic, we hold that the value of Bitcoin may be expressed as a combination of its art and use values as in the following formula:[4]

$$Bitcoin\ value = (Art\ value) * (Use\ value) \qquad (1)$$

Formula (1) shows that the Bitcion value is composed by two arguments, one is its art value and the other is its use value. The Bitcoin value increases in both values, but will shrink to zero if either one goes to zero.

The use value of Bitcoin is easy to understand, and it can be measured by the number of Bitcoin users and the frequency of their usage in reality. Then what exactly is the art value of Bitcoin? Literally, it means that we shall evaluate Bitcoin as if we are evaluating an artwork, which generally attaches little value to the forgery. The following will expand our explanation of this theory, starting from the art value of Bitcoin.

### 3. Art value of Bitcoin

Moving away from the traditional polarized settlement mechanism, Bitcoin designs a purely decentralized scheme. In order to form trust among users, the code of Bitcoin platform must be open-sourced, implying that the underlying code is free for anyone to view, inspect, and use. This in turn facilitates replications as the cost of copying is low and no entity has the duty to punish such behaviors. It, of course, is the side effect of decentralization.

One of the earliest replicas of Bitcoin is Litecoin, which came into being just two years after the emergence of Bitcion. It modifies Bitcoin's mining algorithm to avoid the concentration of mining power. Even though Litecoin mimicked Bitcoin, it was still created as an independent new platform. So it did not automatically inherit the users of Bitcoin and had to acquire its own user base from the very beginning.

The doctrine of platform economics shows that in platform competition it is very hard for a new comer to challenge the incumbent. This is because the strong network effect of the incumbent platform effectively prevents users from switching.[5] Litecoin lived through Bitcoin's nascent stage when its network effect was still relatively weak. But nowadays it is almost impossible for a new coin like Litecoin to survive from the competition of dominant cryptocurrencies. Therefore, such simple imitation as Litecoin (forming a new platform) cannot constitute a real challenge to Bitcoin, until the wide adoption of a totally different technology—bifurcation.

---

[4] Note that other functional forms may also be suggested, which is subject to the future empirical investigations. But the qualitative relationship between the art and use values here will not be affected.
[5] For example, even if people can set up another social media platform like facebook, but it is very hard to attract new users. The user network already formed in facebook greatly increases a user's adhesiveness hence his switching cost.

3.1 Bifurcation

Bifurcation is also called hard fork, which is a technique that splits the old blockchain of a cryptocurrency and generates a new chain permanently as depicted in Figure 1. A contentious bifurcation is primarily due to the disagreement within the current users on the platform protocol. What makes a bifurcation different from creating a new coin is that a bifurcation can effectively circumvent the old platform's network effect, where Bitcoin Cash (BCH) was a prominent example.

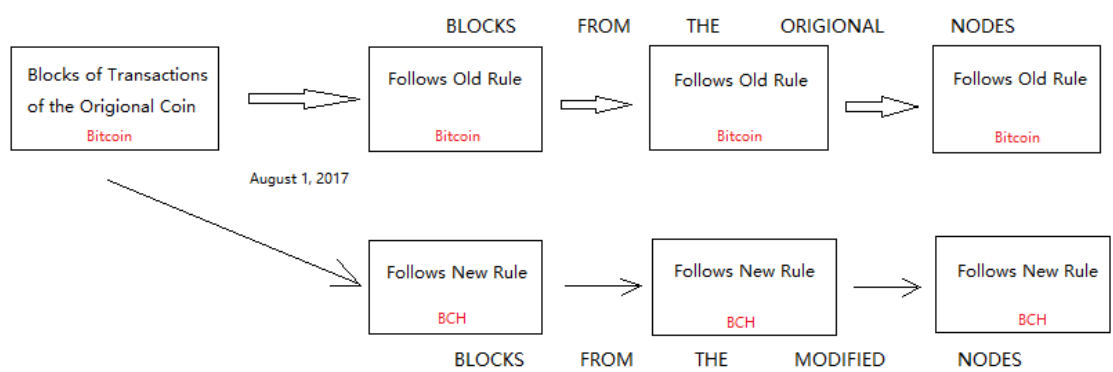

**Figure 1. Blockchain Bifurcation (From Bitcoin to Bitcoin and BCH)**

BCH shared the same blockchain ledger with Bitcoin before August 1, 2017 and, thus, also its users and the related wallet addresses. On August 1, BCH separated from Bitcoin and entered into a new blockchain ledger of its own. Everyone who had bitcoins before the bifurcation received the same amount in BCH credited in their wallets. The new coin, BCH, was able to implement this "free candy" incentive scheme because BCH shared the same wallet addresses with Bitcoin until the bifurcation.

Giving away BCH to the original Bitcoin users entails zero cost from the algorithm. More importantly, these free candies attract the users to stay and transact on the BCH platform. So even being a new platform, BCH avoids the immense cost associated with obtaining users. That is why such practice as bifurcation becomes so popular.

For any cryptocurrency, as long as it becomes successful and reaches a relatively high market value, it always incentivizes the potential bifurcation from itself. For example, BCH itself was further bifurcated into Bitcoin SV on November 16, 2018.

Bifurcation can greatly damage a crytocurrency's market position. Tu and Xue (2018) study the effect of the bifurcation of Bitcoin on its interactions with its substitute, Litecoin. The empirical results show that return and volatility spillovers run in one direction only, i.e., from Bitcoin to Litecoin, before Bitcoin's bifurcation by BCH, while with the direction of shock transmission being reversed after the bifurcation. Obviously, the bifurcation has markedly weakened the pricing influence of Bitcoin within the cryptocurrency market.

As the first serious bifurcation from Bitcoin, BCH's ambition is to overtake Bitcion. It advertises its higher efficiency of settlement by enlarging the original block size from 1M to 8M. However, the grave logical loophole in their argument is why 8M. Can it be more efficient if the block size be further enlarged to 16M, 32M,... etc.？ Indeed, Bitcoin SV bifurcated from BCH by claiming to raise its block size to 1T!

The possibility of the infinite bifurcations of any cryptocurrency implies the potential supply of infinitely many new coins in the market. This dreadful outlook eventually destroys the investors' confidence in cryptocurrencies. We hold that the recent market crash in November, 2018 may just be triggered by the BCH's own bifurcation.

Since there is virtually no barrier to stall a bifurcation for any cryptocurrency, the bifurcation poses a persistent threat to the cryptocurrency market as a whole. It is crucial to find new theories to determine the intrinsic values of various cryptocurrencies, if these markets are to survive into the future. We argue that Bitcoin's intrinsic value can hinge on its art value, because the market for artworks long faces similar replication problems, while they find the solution by only valuing the original copy.

3.2 Art value

The value of an artwork is only stored in its original copy. Any forgery, identical or even better, loses value. Newman and Bloom (2012) propose two psychological mechanisms underlying the special value of original artwork: the assessment of the art object as a unique creative act (performance effect) and the degree of physical contact with the original artist (contagion effect). These two effects are supported by the corresponding experiments in their study. Once the Bitcoin can be treated as a special kind of artwork, we can argue that its value will be stored only in its original form of algorithm.

The next critical question is for an original artwork how to assess its art value. Based on the art literature (e.g. Lazzaro, 2006; Galenson, 2009), we propose that an artwork obtains the art value mostly from two aspects: one is its creative art technique and the other is its cultural significance. The more innovative and sophisticated the technique is, the higher value the artwork can acquire; the greater cultural significance the artwork carries, the higher value it can obtain.

This framework well fits our analysis of Bitcoin. From the cultural perspective, Bitcoin appeared in 2009, a historical moment when the global economy was in crisis while the government intervention by the central bank was highly controversial at the time. Bitcoin successfully landed the spirit of *Austrian economics*, and designed a decentralized system as an alternative for the conventional banking. It for the first time rips the role, and consequently the power, of financial intermediary in a monetary exchange system.

From the technical perspective, Bitcoin invented a combination of technologies to make a decentralized settlement scheme possible. These technologies find many other applications and trigger a lot of innovations in other areas as well. Put together, we hold that Bitcoin deserves the status of an important artwork in human history.

In addition, besides Nakamoto's first message in the genesis block of Bitcoin, experts spot such cultural artifacts as covered-up love messages, poems and ASCII drawings, captions to pictures and eulogies, which are forever built into Bitcoin's blockchain. On the contrary, the identity of the creator of Bitcion, Nakamoto, is still unknown. All these make Bitcoin a unique performance art to some extent.

Only when the market forms such consensus that Bitcoin's value is stored in its art value dimension, can Bitcion get rid of the threat of bifurcation once and for all. After all, unlike the mass-production goods, any recreation of an artwork will not damage the value of the original copy.

Moreover, those rules for the appraisal of artworks can be readily extended to guide our assessment of the values of Bitcoin as well as other bizarre cryptocurrencies.

## 4. Use value of Bitcoin

An artwork can be used either for the personal enjoyment or for the social event. For whatever purpose, it must be accessed in some way rather than being completely locked in the safe. Although the value of an artwork stores in its original copy, its image can always be spread and reproduced. The more an artwork is studied, exhibited and reprinted, the more likely its value is recognized and appreciated, and the higher price it can achieve. Similarly, Bitcoin's value is also scaled over its art value by its use value.

As Bitcoin forms a network, its use value may also follow Metcalfe's law (e.g. Peterson, 2018). It says that the value of a network is proportional to the square of the number of users on the network. For example, the value of facebook may increase exponentially with the number of users registered. Therefore, we suggest that Bitcoin's use value can be measured by both the unique active addresses per day and the number of transactions per day.

From Figure 2, we can see that since year 2017, the growth of both Bitcoin's active addresses and the number of transactions per day started to slow down. On the contrary, the Bitcoin price began a hectic rise, which couldn't be justified by its use value increase. At the same time, its art value was obfuscated by a series of bifurcations as stated above. They together may explain the burst of Bitcoin bubble in 2018.

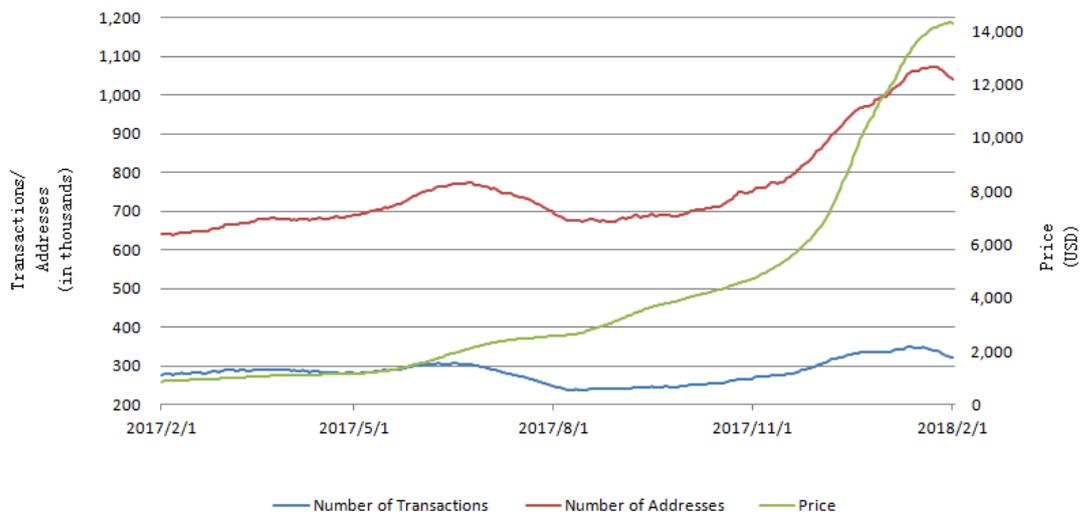

**Figure 2. Prices, active addresses and number of transactions per day for Bitcoin.**

Note: The data are drawn from coinmetrics.io. To smooth the daily fluctuations, we plot the 60-day moving averages for all three indicators in Figure 2.

Obviously, the use value is another important dimension of Bitcoin value. In order to investigate the nature of its use value, we need understand first why the Bitcoin's usage growth became stagnant after 2017. We think the main reason is probably due to its rapidly increasing transaction fees and prolonged transaction confirmation time.

As more and more people begin using Bitcoin, the continuously increasing transaction volume eventually meets the constraint of its block size design, which is set as 1M in the original Bitcoin protocol. In other words, only 1M of data is allowed every 10 minutes. Then more transactions will have to pile up to wait for the confirmation, which may take hours or even days to complete. The competition for the early confirmation eventually bids up the transaction fee, which can be as high as $55 per transaction on average at the end of 2017.[6]

Such a high transaction cost makes it unrealistic for Bitcion to be used in everyday life, such as buying a coffee or a meal. The maximum transaction volume that Bitcoin platform can support each day is largely determined by its original design of 1M block size. This implies that Bitcoin's use value may face a glass ceiling.

After some unsuccessful attempts of improvements on this problem, the market starts turning for other altcoins as a solution. This is exactly the motivation for Bitcoin Cash. BCH has the same code as Bitcoin except that it enlarges the block size to 8M. So in general, the speed of confirmation of BCH will be eight times faster than that of Bitcoin, and it also has much lower transaction fees.

Why does Bitcoin insist on the IM block size without actually updating the original algorithm written by Nakamoto almost a decade ago? The reason is that altering the original code of Bitcoin amounts to modifying the original copy of an artwork. Even if the original artwork may have certain drawbacks, changing it is the same as destroying it. To preserve its art value, the original Bitcoin code needs to be kept intact.

---

[6] See bitinfocharts.com. The transaction fee also dropped after the Bitcoin price crashed in 2018.

However, we can't ignore that the use value is also a vital component of Bitcoin value. To boost its use value, the code needs to be modified to cope with the increasing demand of transactions. Here we clearly see the conflict of Bitcoin's art and use values. How can we resolve this dilemma then?

## 5. Conflict of art and use values

As discussed above, the art value of an artwork is measured by its innovative technique and its particular cultural significance at the time of its creation. After ten years' scrutiny by the market, we think that Bitcoin's art value may have been largely factored into its price. Currently, what hinders the growth of Bitcoin price is its use value. Even though more and more people are willing to accept Bitcoin, its use value is largely limited by its original design of the block size.

5.1 The dilemma between the art and use values

As we argued, Bitcoin cannot increase its use value by simply altering its code if further considering the effect on its art value component. Modifying an artwork is like changing the archive, which destroys people's memory of its cultural significance at the particular historical moment. If we keep updating Bitcoin's algorithm to meet the new demand, we then treat Bitcoin as a pure consumption product, such as the computer operating system.

In a competitive market, a consumption product is priced at its marginal cost. In the decentralized world of cryptocurrencies, people can always copy an open-sourced platform for free, while at the same time use the bifurcation technology to bypass the moat of the network effect of the old platform. This means that the marginal cost of reproduction of a cryptocurrency platform is zero. This explains why in formula (1) when the art value goes to zero, the total value will shrink to zero as a consequence.

To resolve this dilemma, we may draw the lesson from the art industry. How do people balance the art and use values there? On the one hand, an important artwork must be stored in its original form and kept away from any damage. On the other hand, it can be exhibited with heightened security, and its image is authorized to be reproduced or even altered; its significance is discussed and studied. When more people understand and appreciate the artwork, the higher value it may eventually achieve. No one could improve the value of an artwork by updating it with some modern notion or newly developed technique.

So the idea from the art industry is to keep the original copy intact, while encourage the spread of the image via various means, e.g., exhibition, conference, seminar, publication and TV program etc. We believe Bitcoin could borrow this approach to break the ceiling of its use value without sacrificing its art value.

5.2 A Bitcoin-Mirror proposal

Following the above logic, we propose an image coin called Bitcoin Mirror, or BTCM, as a solution to the dilemma that Bitcoin faces.

BTCM is an independent new cryptocurrency platform like Litecoin, not a bifurcated platform derived from Bitcoin like BCH. However, BTCM is pegged with Bitcoin with a predetermined exchange rate. The issue of BTCM must be based on the collateral of corresponding amount of Bitcoin, and BTCM and Bitcoin can be exchanged freely via a decentralized and open-sourced blockchain exchange.[7] This exchange acts as a decentralized digital central bank that adjusts the supply of BTCM based on the collateral of Bitcoin that it retrieves from the market. The mechanism of BTCM is depicted in Figure 3.

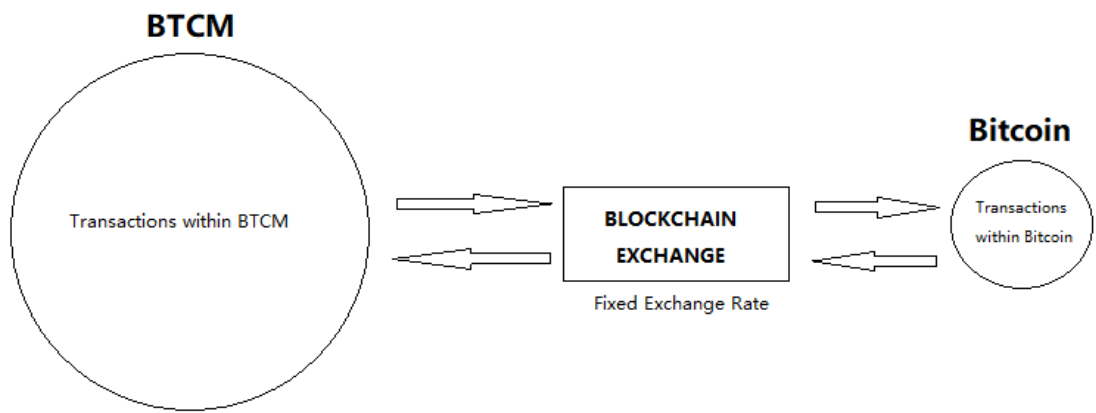

Figure 3. The Mechanism of the Bitcoin Mirror (BTCM)

The scheme is like gold standard during Britten Woods period with just a vital difference: unlike Fed, this decentralized blockchain central bank will never deviate from the consensus code. As the image of Bitcoin, BTCM can be given great flexibility to cope with any future situation. Its block size can be set much higher than that of BCH for a much speedier confirmation, making it comparable to PayPal. Continuous evolution of BTCM's algorithm can support much larger volume of transactions, helping it truly integrate into our daily life. We think that in the future probably only the large block trading may remain within Bitcoin.

Following the above design, the further bifurcation of BTCM can also be prevented because the bifurcated coin cannot provide free candy to lure the users to transact on the new platform any more. It needs the corresponding Bitcoin as the collateral! Moreover, the emergence of other competing image coins will not cause a problem either. They only render the market more choices, but cannot change the monetary supply as their issuance must be linked to Bitcoin.

---

[7] The technical feasibility of such an exchange can be further discussed. It may need an intermediary currency as the exchange media.

Hence the bottom line is that Bitcoin's original algorithm must not be altered. Its use value shall be promoted by an independent new image coin platform. If successful, it will completely lift the upper bound of the usage constraint for Bitcoin. Other improvements that need to connect to the Bitcoin code will always fall into the trap of art and use values dilemma. So we believe that BTCM may be an ideal solution to this problem.

## 6. Application of the theory to other cryptocurrencies

The above illustrates the normative dual-value theory of Bitcoin. This theory has more general applications to the other cryptocurrencies as well, because they all share a common feature: the conflict between art and use values.

During the cryptocurrency's boom market in 2017, large number of Initial Coin Offerings (ICO) poured into various exchanges every day. Driven by the explosive listing demand, lots of new exchanges also opened in a short time. Every coin self-claimed that it was a revolution. But how to judge the true value of each coin was a hard task as no assessment guideline was agreed upon.

We hold that the dual-value theory can provide an effective guidance for our evaluation of a new cryptocurrency. The following analysis applies this theory to several major coins in the market, such as BCH, ETH and EOS.

5.1 BCH

Bitcoin Cash, or BCH, claims to improve on Bitcoin's confirmation algorithm so that the transaction cost can be largely reduced. However, BCH achieved this goal via a bifurcation from Bitcoin with a simple modification of the block size. Although its potential use value may be increased due to lower transaction cost, it mostly loses the art value.

If we compare Bitcoin to the famous painting *Mona Lisa* by Leonardo da Vinci, then BCH is analogous to *Prado Mona Lisa* of an unknown painter as in Figure 4. It was reputed to be the earliest replica of Da Vinci's Mona Lisa with a sharper contrast of colors, which may be more eye-catching for viewers. As a replica, *Prado Mona Lisa* has been considered worthless due to the lack of art value for centuries.[8] In contrast, the unique technical as well as cultural significance of the renaissance art period is stored in the original copy of Mona Lisa, making the original one a priceless artwork.

BCH reached 4000 dollars shortly after its launch. The price was mostly driven by the speculative capital backing it. The logic of the capital was to quickly accumulate more users than Bitcoin and surpassed it via the platform network effect. This is the common competition strategy in platform industry, normally leaving the game a pure capital war.

However, the situation is a little different in the world of cryptocurrencies because of the possibility of bifurcation. Only relying on the network effect cannot constitute a moat for an open-sourced cryptocurrency platform as it can be costlessly circumvented by bifurcation. In fact, the true moat lies in the cryptocurrency's art value.

---

[8] Some recent findings about Prado Mona Lisa may affect the previous assessment of its value, but it's not relevant to our analogous discussion here.

Without a solid anchor of art value, the market value of BCH can be easily watered down by any subsequent bifurcation from itself. Indeed, BCH did not wait long to see its own bifurcation in November, 2018. Ironically, its derivative Bitcoin SV picked up BCH's old claim that the block size was still too small and need further enlargement. By formula (1), any reproduction that ignores the art value and only targets on the use value will receive a zero total value in the long time. Both BCH and Bitcoin SV may eventually step into this path.

5.2 ETH

ETH is currently the second largest cryptocurrency in terms of market capitalization. ETH is the coin to support Ethereum, an open-sourced blockchain platform that runs the smart contracts. The founder of Ethereum is Vitalik Buterin, who identifies Bitcoin's idea of decentralization of confirmation but regrets its lack of extendibility (as it can only be used for the monetary exchange). Therefore, Ethereum aims to break the application boundary of Bitcoin, and try to support any type of exchange activities in the form of smart contracts.

Obviously, ETH's art value derives from Ethereum's breakaway from Bitcoin's confinement in the monetary transactions. We can compare it to *Les Demoiselles d'Avignon* by Picasso, which disrupts the rules of classical paintings and changes our way to view the world. Consequently, it merits a high art value.

Notice that ETH is actually a bifurcation from ETC, the earliest coin to support Ethereum platform, in July 2016. Then how does the dual-value theory explain the surprising fact that ETH's market capitalization eventually becomes more than thirty times that of ETC?[9] Why didn't ETH sacrifice its art value when updating from ETC?

First of all, the key difference between ETH and Bitcoin is that the former is still an on-going work while the latter is a finished one. Bitcoin's complete and also rigid structure on currency transactions provides a clear scope for the art value assessment. In contrast, Vitalik Buterin, the founder of Ethereum, claimed that Ethereum was still an evolving experiment, which amounts to that the painter thinks the work is not done yet.

Second, the bifurcation of ETH from ETC was brought about by the "painter" Buterin himself, naturally representing an improvement of an artwork in progress. It did not lose the art value because it is still connected with the original author (referring to the performance and contagion effects).

Furthermore, the bifurcation of ETH was not motivated by the personal or business interests. Rather, it aimed to return the hacked coins of an organization running in Ethereum. By doing so, it won more users. Eventually, ETH dominates ETC in both art and use values.

---

[9] Calculated by the data from CoinMarketCap.

The biggest caveat for ETH lies exactly in the uncertain nature of the platform since the "painter" Buterin is still exploring. It has both the potential to surpass Bitcoin and the risk to collapse to worthless. Fortunately, the market has a relatively stabilized ETC as a hedge to ETH's downside risk to some extent. So overall, we hold quite a positive expectation for ETH.

5.3 EOS

EOS platform does not revolutionize on Ethereum, rather it also claims to be a decentralized operating system for "everything". What is new is that it upgrades to support industrial-scale decentralized applications. It was created in June, 2017 and sponsored by a company. The market capitalization of the platform coin, also called EOS, quickly reached about a quarter of that of Bitcoin.

The objective of EOS is clear at the beginning, to increase its use value. Instead of verifying the state of the network at any given time (like Bitcoin or ETH), EOS nodes only verify the series of events that have occurred so far to keep track of network state. So the system claims to be able to scale to one million transactions per second out of the gate on a single machine.

Even though EOS adopts the blockchain technology, it is much more polarized than Bitcoin or ETH, as it can only have 21 block producers. In order to provide convenience to the development of the decentralized applications on the platform, EOS also provides services like user authentication, cloud storage, and server hosting like other traditional business entities.

Clearly, EOS moves away from the spirits of Bitcoin and ETH. It sacrifices the degree of decentralization for much more efficient operation environment. Relatively speaking, EOS has far closer connection with the traditional business models; hence it may obtain much less art value.

Using the analogy of artworks, EOS can be compared to the collage, e.g., the *Indian Dancer* by Hannah Hoch, because it mixes the abstract painting with other concrete daily articles.[10] A collage is generally priced lower than a painting in the art market because it is often seen as a mixture of "high" and "low" art.[11]

Like ETH, EOS is still evolving and even riskier. Its total value will be largely determined by the use value it can finally accomplish. In a sense, EOS may have to compete in the fields consisted of other more traditional tech companies in the future.

---

[10] For a consistent comparison with other paintings, we select a collage with a facial image by Hannah Hoch. She is renowned for her invention of photomontage, a type of collage in which the pasted items are actual photographs, or photographic reproductions pulled from the press and other widely produced media.
[11] The high art means the traditional definition of fine art and the low art refers to that made for mass-production or advertisement.

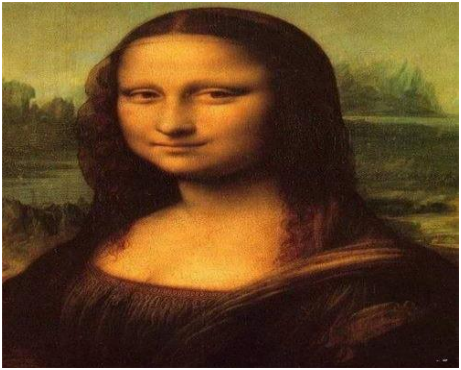
***Bitcoin*** / *Mona Lisa*

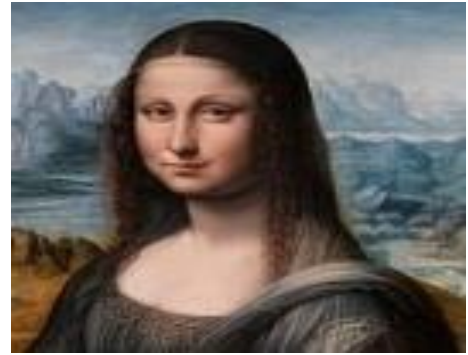
***BCH*** / *Prado Mona Lisa*

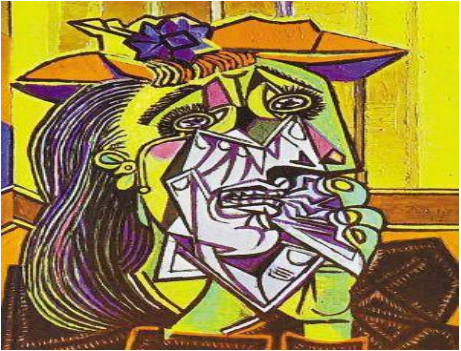
***ETH*** / *Les Demoiselles d'Avignon*

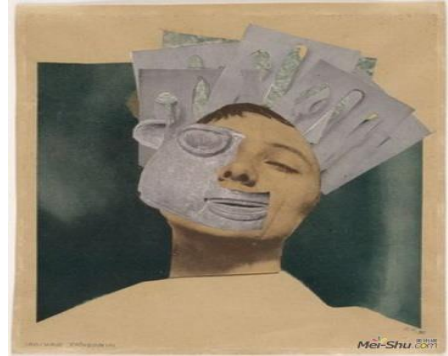
***EOS*** / *Indian Dancer*

**Figure 4. Analogous paintings for four major cryptocurrencies**

To summarize the above analysis, we can produce the ranking of the art values of four major cryptocurrencies as:

$$Bitcoin>ETH>EOS>BCH \qquad (2)$$

and the ranking of the use values of them as:

$$EOS>ETH>BCH>Bitcoin \qquad (3)$$

Bitcoin has the highest art value but lowest use value because it is constrained by its original code on the block size. We propose a Bitcoin Mirror (BTCM) scheme that can relieve this kind of inherent limitation. If it can be successfully realized, we believe the Bitcoin value may rise to an unthinkable level.

## 7. Conclusion

We observe that Bitcoin as well as other cryptocurrencies are all plagued by the impact from bifurcation. A coin's market position can be continuously weakened by successive bifurcations.

Since the marginal cost of bifurcation is theoretically zero, it inevitably causes the coin holders to doubt on the existence of the coin's intrinsic value. If no constructive consensus can be reached on this point, the outlook of cryptocurrencies is dim.

This paper suggests a normative dual-value theory to assess the fundamental value of Bitcoin. We draw on the experience from the art market, where similar replication problems are prevalent. The idea is to decompose the total value of a cryptocurrency into two parts: one is its art value and the other is its use value. Such a construction responds to both the bifurcation and application challenges that a coin faces. The tradeoff between these two values is also analyzed, which enlightens our proposal of an image coin for Bitcoin so as to elevate its use value without sacrificing its art value.

We apply the dual-value theory to evaluate the prospects of four major cryptocurrencies based on their functional designs: Bitcoin, BCH, ETH and EOS. Our theory can be readily extended to other cryptocurrencies as well. We find this framework is helpful for both the investors and the exchanges to examine a new coin's value when it makes a first appearance in the market. The appropriate empirical tests for this theory may be left for our future studies.